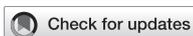

# What is *ab initio* in nuclear theory?

A. Ekström[1]\*, C. Forssén[1], G. Hagen[2,3], G. R. Jansen[2,4], W. Jiang[1] and T. Papenbrock[2,3]

[1]Department of Physics, Chalmers University of Technology, Göteborg, Sweden, [2]Physics Division, Oak Ridge National Laboratory, Oak Ridge, TN, United States, [3]Department of Physics and Astronomy, University of Tennessee, Knoxville, TN, United States, [4]National Center for Computational Sciences, Oak Ridge National Laboratory, Oak Ridge, TN, United States

*Ab initio* has been used as a label in nuclear theory for over two decades. Its meaning has evolved and broadened over the years. We present our interpretation, briefly review its historical use, and discuss its present-day relation to theoretical uncertainty quantification.

## 1 Introduction

The literal meaning of the latin term *ab initio* implies that one starts from the beginning. In computations of atomic nuclei, this means that the relevant degrees of freedom should be quarks and gluons. However, the history of physics tells us that we do not need to know everything to describe something and that we have some freedom in choosing the starting point. As such, we do not necessarily have to employ Standard Model degrees of freedom. In fact, many nuclear properties were successfully analyzed in terms of hadronic degrees of freedom before we even knew about the existence of quarks [1–3]. Today, we know how to explain this using renormalization group (RG) ideas [4, 5]. One may wonder about the exact meaning of the *ab initio* method and what should constitute the beginning. However, it is safe to say that a hallmark of this approach is its promise of precise and accurate predictions, with quantified uncertainties, across the multiple energy scales relevant to nuclei. Examples range from low-energy collective phenomena such as deformation and rotation [6–10], to loosely bound and unbound nuclei [11–16], and to lepton nucleus scattering in the quasi-elastic energy regime [17–19]. We expect the *ab initio* method to reliably extrapolate, in a controlled and systematic way, to regions outside the ones used for inferring the model parameters. Following the ideas from effective field theory (EFT) [20], *we interpret the ab initio method to be a systematically improvable approach for quantitatively describing nuclei using the finest resolution scale possible while maximizing its predictive capabilities.* A key part of this interpretation is the possible tension between the two latter aspects. In a nuclear physics context, we therefore let nucleons, and possibly other relevant hadronic degrees of freedom, define the beginning. Lattice quantum chromodynamics (QCD) might one day be the optimal starting point for predicting nuclear phenomena. Presently, Lattice QCD continues to provide useful input for EFTs based on hadronic degrees of freedom. However, it currently lacks predictive power for describing atomic nuclei [21–25].

We acknowledge that the *ab initio* method is interpreted differently by different people; see, e.g., Refs. [26–33]. In nuclear physics, the evolution of *ab initio* and its wide application reflect the creativity and innovation of the scientists who perform *ab initio* computations. In this sense, *ab initio* is unlike Tennessee Whiskey or French Champagne, which are internationally protected labels, but rather like Gruyère cheese, i.e., a generic expression that benefits a "vibrant, competitive marketplace"[1]. In this review we provide a brief history of *ab initio* nuclear physics (Section 2), clarify our interpretation of this label (Section 3), explain how this approach creates an inferential advantage (Section 4), and provide examples in connection with some remaining challenges (Section 5).

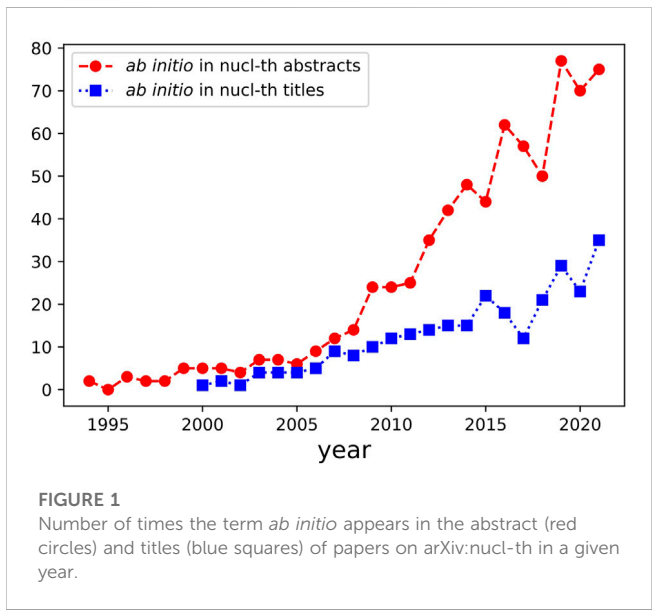

FIGURE 1
Number of times the term *ab initio* appears in the abstract (red circles) and titles (blue squares) of papers on arXiv:nucl-th in a given year.

## 2 A brief history of *ab initio* nuclear physics

A search of the term *ab initio* in the title on arXiv:nucl-th returns about 300 papers, with the earliest one by Navrátil, Vary, and Barrett (Ref. [34]) dating back to the year 2000. When the search includes abstracts, the count increases to more than 700, and papers by Leinweber (Ref. [35]) and Friar (Ref. [36]) are the earliest published in the mid-1990s. Since then, an ever-increasing number of authors have used the term *ab initio* to characterize their work. In Figure 1, we show a plot of the data for the yearly use of this term in titles and abstracts.

The authors of Ref. [34] did not explain what distinguished their *ab initio* no-core shell model computations from quite similar earlier approaches [37, 38] (see also Ref. [39]). Whatever the reason, the term *ab initio* stuck and has been popular ever since. Colloquially, we often use *ab initio* to label theoretical analyses of nuclei based on "realistic" nucleon-nucleon, and three-nucleon potentials, with solutions to the nuclear many-body problem obtained either "virtually" exactly or with controlled approximations. Over the years, however, the small number of available "realistic" or "high-precision" nucleon-nucleon potentials [40, 41] have been replaced by nucleon-nucleon potentials *plus* three-nucleon potentials from chiral effective field theories ($\chi$EFTs) of QCD [42–44].

---

1 https://www.nytimes.com/2022/01/12/business/gruyere-cheese-us-court-ruling.html

Because of the power counting in $\chi$EFT the potentials are recognized as approximate with a fidelity that presumably increases with increasing chiral order. This presented an opportunity for systematically improvable many-body methods that scale polynomially with increasing mass number [26, 30, 45–52]. Why solve an approximate potential virtually exactly? This class of gently-scaling methods has now extended the reach of many-body calculations to medium-mass and heavy-mass nuclei [13, 53–58]. The computational cost of these calculations is kept manageable by also approximating three-nucleon potentials as normal-ordered, i.e., "density-dependent," two-body potentials [53, 57, 59, 60], and using the intrinsic kinetic energy alleviated problems with the center of mass in the laboratory system [61–63]. These efforts also revealed the need for nuclear potentials that accurately reproduce bulk observables beyond the lightest-mass nuclei [64, 65]. This spurred the development of many new potentials differing by the degrees of freedom they used, how the numerical values of the low-energy constants (LECs) were determined, the choice of regulator function, power counting, and degree of locality [64–79].

Since the mid-to-late 1990s, this two-decades-long struggle to describe nuclei has brought nuclear structure and reactions closer together [11, 12, 80, 81]. Ideas from EFT [82] and RG [83, 84] have changed our views on what is observable [85], the importance of understanding the intrinsic resolution-scale and scheme dependencies [86, 87], how we can systematically account for finite-size corrections [88, 89], and estimate the effects of truncating the EFT expansion [69, 72, 90, 91]. These ideas have also led to the advent of the in-medium similarity renormalization group [49, 51, 52] and nuclear lattice EFT [92, 93] as the latest many-body methods. What we nowadays refer to as *ab initio* computations of nuclei [16, 56, 58, 94] is intimately linked to the ideas of EFT and uncertainty quantification. Clearly, what we considered *ab initio* two decades ago does not necessarily pass as *ab initio* today, and *vice versa*.

## 3 Our interpretation of the *ab initio* method

The methods of EFT [82] and RG [83, 84] provide a valuable foundation for the idea that the physics at a given energy scale does not explicitly depend on the details at much higher energies. The beginning can therefore be marked by identifying a scale separation, specifying the relevant degrees of freedom and symmetries, and allowing interactions accordingly. A power counting facilitates meaningful truncations.

We interpret the *ab initio* method as employing Lagrangians, Hamiltonians, or energy density functionals based on EFT principles and with degrees of freedom chosen such that it maximizes our predictive capabilities. *Ab initio* descriptions of atomic nuclei concern the physics of multi-hadron systems in an energy range from keV to a few hundreds of MeV. As such, it is reasonable to start from hadronic degrees of freedom with interactions derived from the Standard Model using the principles of EFT. While hadrons are composite systems, and QCD is the underlying theory of the strong nuclear force, Lattice QCD calculations of two-hadron systems are not yet under control [21, 22]. This might change, and one could imagine computing nuclei *ab initio* from QCD. Moving *the beginning* from hadronic degrees of freedom to quarks and gluons would extend the upper limit of the applicable energy scale by several orders of magnitude and thus increase predictive capabilities significantly. It is, however, an open question whether this ansatz will capture emergent phenomena like the saturation of nuclear forces [95]. Even if this were possible, it is another question how much understanding would be gained about emergent phenomena that involve novel (low-resolution) degrees of freedom from such a high-resolution perspective. The usefulness of the tower of EFTs will most likely remain [96].

Assuming that *ab initio* descriptions of nuclei inherit the physics of the Standard Model *via* EFT methods, we expect to obtain more reliable predictions compared to complementary and phenomenological approaches. Also, building on an EFT, the *ab initio* method should be systematically improvable, organizing the relevant physics according to importance following the principles of power counting. To use this advantage, we must obtain observables using numerically exact methods or, if necessary, using controlled approximations that allow for a systematic analysis. By controlled approximations, we mean ignoring, in a graded way, what we believe to be less essential physics. Doing so, we obtain a handle on what we discard and a more meaningful estimate of our prediction uncertainty. We would like to emphasize the distinction between ignored physics and unresolved physics. An example of the latter is short-range physics that, although unresolved, is accounted for in the Hamiltonian *via* contact interactions [97].

It is pivotal to incorporate and declare our knowledge base and assumptions in analyzing uncertainties. The *ab initio* method does not emerge from a vacuum. That would be an *ex nihilo* method, of which we cannot find any example in science. Quantifying theoretical uncertainties grounded in systematicity should create an advantage when assessing discrepancies between theory and experiment. Note that according to our interpretation, the *ab initio* method does not guarantee that we can find absolute bounds on the theoretical uncertainties nor that we approach the true data-generating mechanism by gradually reducing all truncations. Indeed, should tensions between experiment and theory remain despite our best efforts to quantify uncertainties and keep the truncations at a minimum, we obtain quantitative evidence that we should contest at least one of our assumptions.

## 4 How the *ab initio* method creates an inferential advantage

The use of probability theory to quantify uncertainty plays a central role in the scientific endeavor of inferring new knowledge about the Universe. In this context, the *ab initio* method has evolved significantly over the last few years and now offers a distinct advantage. However, before we can elaborate on the topic of inductive inference and its relation to the *ab initio* method, we must briefly discuss the nature of science in terms of data, theories, and models. This topic is expanded upon in the context of EFTs much more thoroughly in, e.g., Refs. [98, 99].

Let us start with the data $\mathcal{D}$ obtained through a measurement process. All data are equipped with uncertainties of various origins; let us denote this $\delta\mathcal{D}$. Given some data $\mathcal{D}$, one could ask what this data can tell us about future data $\mathcal{F}$. At present, the future data is uncertain and must therefore be described with a conditional probability $p(\mathcal{F}|\mathcal{D},I)$ [100]. Here $I$ denotes all available background information. The obvious question is: How does one go from this abstract probability to something that can be evaluated quantitatively? The answer is to develop a theory within which we can formulate a model that allows for numerical evaluation.

In physics, a theory is very often some framework that postulates or deduces from some foundational principles the spacetime dependence of a system of interacting bodies, e.g., Einstein's field equations in the general theory of relativity or Heisenberg's equations of motion in quantum mechanics. A physical theory always comes with some prior probability of being wrong and this probability should never be exactly zero or one. Otherwise no new evidence/data will ever influence the validity of the theory. In this sense, all theories are wrong, i.e., never correct with absolute certainty. This provocative statement is designed to draw attention to the fact that all theories can be improved or replaced as we progress and gather more data.

A physical model $M$ allows quantitative evaluation of the system under study. Any model we employ will always depend on model parameters $\boldsymbol{\theta}$ with uncertain numerical values. Moreover, like theories: "all models are wrong" [101]. Indeed, there will always be some physics that we still need to include or are unaware of today. If we denote the mismatch between model predictions and data as $\delta M$, we can write

$$\mathcal{D} = M(\boldsymbol{\theta}) + \delta\mathcal{D} + \delta M. \tag{1}$$

We often refer to the mismatch term $\delta M$ as the model discrepancy [102]. Naturally, we are uncertain about this term, so we represent it by a probability distribution following our beliefs about the limitations of $M$. It is no trivial task to incorporate model discrepancies in the analysis of scientific models and data. Nevertheless, it is crucial to avoid overfitting the model parameters $\boldsymbol{\theta}$ and making overly confident model predictions [103]. It is in this context that the *ab initio* method creates an inferential advantage. The promise of systematicity grounded in EFT, and the controlled approximations underlying the computation of nuclear observables, allows us to be quantitative about the distribution that governs $\delta M$ as we increase the fidelity of $M$. For simplicity we sometimes refer to EFTs as models. However, an EFT is more than a physical model in the traditional sense. Indeed, within its domain of applicability an EFT prediction reflects the underlying theory up to a truncation error. In this sense, the EFT is complete, which is a distinct advantage compared to traditional models. This is sometimes referred to as model-independence. Of course, the underlying theory might be wrong, and such model discrepancies cannot be remedied at the level of the EFT.

For example, assume that we operate with an EFT of QCD to derive the potential for the nuclear interaction up to some order in the relevant power counting. In addition, suppose that we use a systematically improvable many-body method at some well-defined truncation level to solve the many-body Schrödinger equation for the ground-state energy in our favorite nucleus. Then we can say more about $\delta M$ than if we use a shell model description grounded in phenomenologically defined interaction matrix elements tailored to a specific model space. We are not saying that the latter calculation cannot provide valuable guidance or insight. However, we are saying that it is possible to systematically test the underlying assumptions within the *ab initio* method. Having quantified $\delta M$ also tells us the significance of a possible discrepancy or tension between experiment and theory.

The distribution of future data conditioned on past data and background information $I$, i.e., $p(\mathcal{F}|\mathcal{D}, I)$, is referred to as a posterior predictive distribution (PPD). Assuming that we have a model $M(\theta)$ for the data-generating mechanism, then we can express the PPD by marginalizing over the uncertain model parameters $\theta$ belonging to some parameter space $\Omega$

$$p(\mathcal{F}|\mathcal{D}, I) = \int_\Omega p(\mathcal{F}|\theta, I) p(\theta|\mathcal{D}, I) \, d\theta. \quad (2)$$

By performing this integral, we average all predictions with respect to our uncertainty of the model parameters $\theta$. To evaluate the posterior probability density function (PDF) $p(\theta|\mathcal{D}, I)$ for the model parameters, we can employ Bayes' theorem, i.e.,

$$p(\theta|\mathcal{D}, I) = \frac{p(\mathcal{D}|\theta, I) p(\theta|I)}{p(\mathcal{D}|I)}. \quad (3)$$

This requires a likelihood function $p(\mathcal{D}|\theta, I)$ and a prior distribution of the model parameters $p(\theta|I)$. The denominator, $p(\mathcal{D}|I)$, does not explicitly depend on $\theta$ and is only needed for proper normalization. Quantifying the posterior is called Bayesian parameter estimation and is a staple of Bayesian inference. It is a probabilistic generalization of parameter optimization and maximum likelihood estimation.

In the historical developments of "high-precision" nucleon-nucleon potentials, one often employed a $\chi^2$-measure to quantify the goodness of fit to nucleon-nucleon scattering data [40, 41]. Although such an approach has several drawbacks, most notably its limited use in uncertainty quantification and non-trivial incorporation of prior knowledge and model discrepancy, it is undoubtedly less demanding computationally than quantifying a multi-dimensional posterior PDF. Evaluating the posterior requires numerical methods like Markov Chain Monte Carlo [104, 105], which is no silver bullet and by no means guaranteed to succeed. To compute the denominator in Eq. (3), i.e., the marginal likelihood, is even more difficult. There is significant progress in linking *ab initio* methods to the Bayesian inferential approach in the nucleon-nucleon and few-nucleon sectors [106]. The development of efficient and accurate emulators [107–110] should provide us with sufficient leverage to continue applying Bayesian methods for analyzing and quantifying uncertainties for non-trivial nuclear structure observables and reaction cross sections. Access to emulators also opens the door to detailed experimental design studies [111–116].

## 5 A few examples

We will briefly discuss a few examples and highlight some remaining challenges to clarify our interpretation of the *ab initio* method to analyze nuclei.

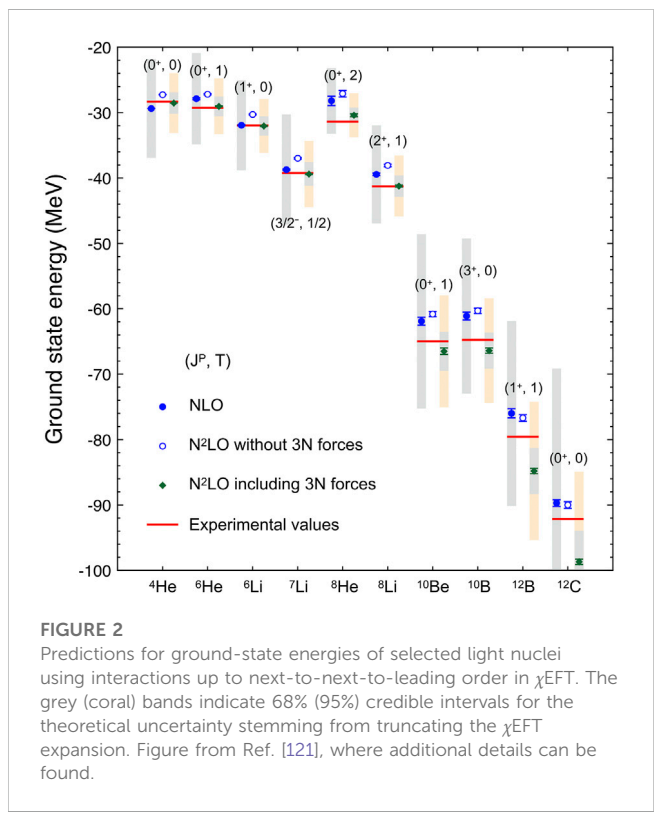

FIGURE 2
Predictions for ground-state energies of selected light nuclei using interactions up to next-to-next-to-leading order in $\chi$EFT. The grey (coral) bands indicate 68% (95%) credible intervals for the theoretical uncertainty stemming from truncating the $\chi$EFT expansion. Figure from Ref. [121], where additional details can be found.

For light-mass nuclei, methods like the Faddeev-Yakubovsky equations [117], hyperspherical harmonics expansion [27], no-core shell model [39], and quantum Monte Carlo [118] yield virtually exact solutions to the many-body Schrödinger equation [119], barring systematic truncations of the single-particle basis and the Hilbert space of many-body wave functions, or limited sampling statistics. As such, the fidelity of the prediction is mainly limited by the available computational resources [120]. When we employ these methods with interactions that can be systematically improved, we obtain the prototypical *ab initio* calculation of a nucleus. As an example, in Figure 2, we show the predictions for ground-state energies in selected nuclei with mass numbers $A = 4$–$12$ as obtained in a systematic study [77] of light nuclei using two-plus three-nucleon interactions up to next-to-next-to-leading order in $\chi$EFT. The parameters of the employed interactions, i.e., the LECs, were calibrated to reproduce selected two- and three-nucleon data. The authors of that study recognized the well-known trend [54] of over-binding starting at $A \approx 10$ and increasing with $A$. Whether going to higher orders in $\chi$EFT ameliorates this issue remains to be understood. A recent paper [122] shows that ground-state energies are better reproduced when going to the next order, but nuclear radii remain challenging to describe.

In contrast to increasing the chiral order of the nuclear potential, which also introduces additional LECs to be inferred, it was demonstrated in Ref. [64] that one could obtain accurate predictions for binding energies and nuclear radii in medium-mass nuclei using a chiral interaction at next-to-next-to-leading order (NNLO$_{\text{sat}}$), which is calibrated to reproduce data for bulk observables in nuclei with A = 2–16, see Figure 3, in addition to nucleon-nucleon scattering cross sections. Calibrating the LECs to reproduce this wider class of nuclear data residing in the domain of applicability of $\chi$EFT, has been very fruitful and informative. This approach was expanded upon in Ref. [65] by exploiting empirical information from nuclear matter at saturation densities and including the $\Delta$(1232)-isobar in the chiral expansion of the nuclear interaction. The strategy of inferring LECs to also reproduce bulk properties of medium-mass nuclei runs the risk of overfitting but there are Bayesian methods to mitigate this, as discussed in the next paragraph. The interactions in Refs. [64, 65] account for rudimentary theory and method errors and can be systematically improved. Therefore we characterize them, and ensuing predictions utilizing controlled approximations, as *ab initio*. On the contrary, nuclear interactions designed to maximize the data likelihood of nucleon-nucleon scattering cross sections at any cost, invoking, e.g., unphysical parameters of the regulator for fitting purposes [78], cannot be considered as an *ab initio* approach despite being based on EFT Lagrangians. A smaller $\chi^2$-value does not imply greater *ab initio* content.

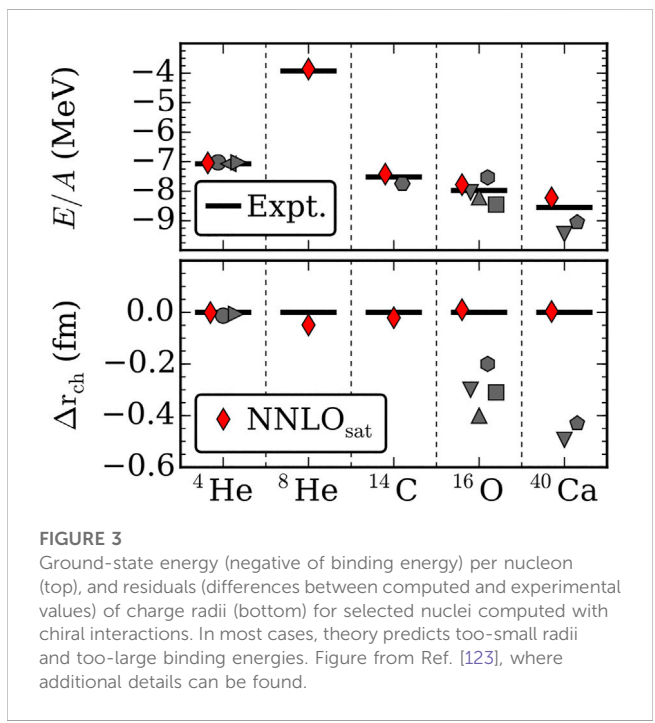

FIGURE 3
Ground-state energy (negative of binding energy) per nucleon (top), and residuals (differences between computed and experimental values) of charge radii (bottom) for selected nuclei computed with chiral interactions. In most cases, theory predicts too-small radii and too-large binding energies. Figure from Ref. [123], where additional details can be found.

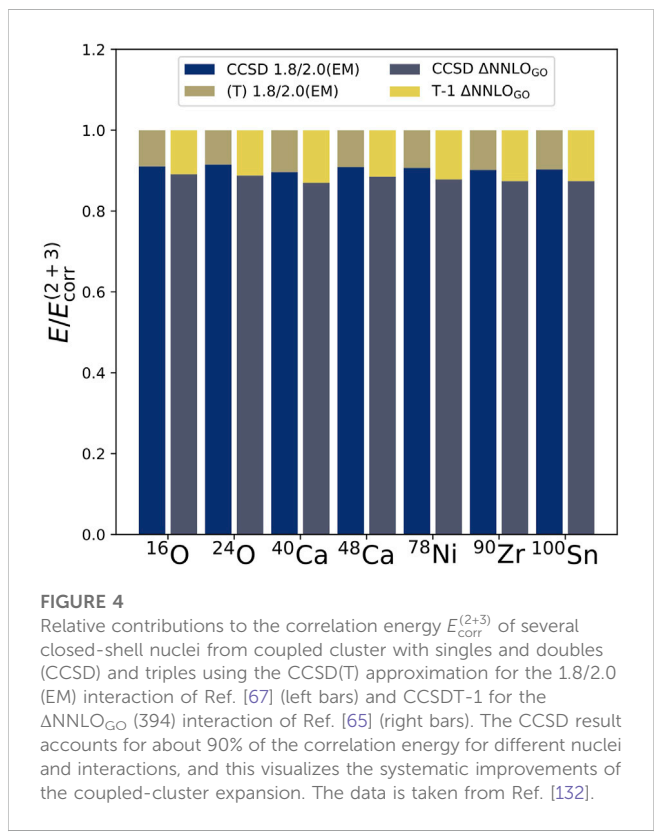

FIGURE 4
Relative contributions to the correlation energy $E_{corr}^{(2+3)}$ of several closed-shell nuclei from coupled cluster with singles and doubles (CCSD) and triples using the CCSD(T) approximation for the 1.8/2.0 (EM) interaction of Ref. [67] (left bars) and CCSDT-1 for the $\Delta$NNLO$_{GO}$ (394) interaction of Ref. [65] (right bars). The CCSD result accounts for about 90% of the correlation energy for different nuclei and interactions, and this visualizes the systematic improvements of the coupled-cluster expansion. The data is taken from Ref. [132].

Utilizing Bayesian inference methods, one can express the PDF for the LECs conditioned on low-energy data while accounting for the truncation errors of $\chi$EFT and our knowledge about the accuracy of the employed many-body methods. The challenge, however, is to quantify these uncertainties. There are methods based on, e.g., Gaussian processes to account for correlated EFT truncation errors in nucleon-nucleon scattering [124] and nuclear matter predictions [125]. However, studies of the $\chi$EFT truncation error for finite nuclei deserve more attention [126]. Hu et al. [58] took the first step in this direction to analyze heavy-mass nuclei, where they quantified an *ab initio* PPD for the neutron-skin thickness in $^{208}$Pb. This distribution is conditioned on low-energy data from light- and medium-mass nuclei together with assigned uncertainties of the employed nuclear interactions and the many-body methods.

To enable predictions for nuclei with $A \gtrsim 10$, one must use methods whose computational complexity scales gently with $A$ and the size of the single-particle basis. Still, these methods must retain essential many-body physics to describe the observable of interest. Three-nucleon interactions can be challenging to handle computationally [127]. Operating with truncated model spaces and normal-order-approximated interactions goes well with the ideas of the *ab initio* method. Here, we exemplify our discussion using the coupled cluster method [30] with polynomial scaling in $A$, but we note that several methods [128, 129] of this kind exist. The coupled cluster method exploits an exponentiated cluster operator $\hat{T}$ expanded on particle-hole excitations of some many-body reference state. Truncating the expansion of $\hat{T}$ at some level of $n$-particle $n$-hole excitations, and solving for the remaining excitation amplitudes, constitutes a systematically improvable description of the many-body wave function. Although a formal bound on the effect of higher-order particle-hole excitations is lacking, it is clear how to improve. This recipe for improvement is what we seek in an *ab initio* method. Unfortunately, the convergence pattern might be irregular and vary significantly depending on the observable considered. For example, a rapid convergence for the ground- and first excited-state energies does not imply that their respective wave functions yield a converged description of non-stationary observables [130]. Convergence must be inspected empirically by either gradually increasing the number of particle-hole excitations, or defining a more appropriate starting point (reference state) for the expansion, see, e.g., Ref. [131]. Understanding this kind of convergence pattern is important for proper uncertainty quantification and remains an open question that requires significant domain knowledge.

In Figure 4, we show how the 2p-2h and 3p-3h excitations of the coupled-cluster method contribute to the correlation energy of various closed-shell nuclei. The correlation energy is defined as the difference between the predicted binding energies and the Hartree-Fock energy. Here, the correlation energy $E_{corr}^{(2+3)}$ is the sum of the 2p-2h correlation energy from CCSD and the 3p-3h correlation energy from the triples, namely the Λ-CCSD(T) approximation [133] (for the 1.8/2.0 (EM) interaction of Ref. [67]) and the CCSDT-1 approximation [134] (for the interaction of Ref. [65]). We see that CCSD accounts for about 90% of the correlation energy for different nuclei and interactions. Various quantum chemistry applications [135] have obtained similar results. This strongly suggests that coupled-cluster theory provides us with a systematic approximation when truncated at the doubles, triples, etc levels. While we do not (yet), have an understanding of the hierarchy (i.e. triples yield much smaller energy contributions than doubles) shown in Figure 4, Sun et al. [132] proposed RG arguments as a possible explanation: Lowering the resolution scale in the three-body subsystems of a many-body system corresponds to removing (short-ranged) triples excitations. Extending the arguments by Lepage [5] and Bogner and Roscher [136] from two to three-body systems then suggests that removal of short-ranged triples excitation can be compensated by a renormalization of the three-body contact.

As discussed, the *ab initio* method aims at maximizing its predictive power over multiple energy scales relevant to nuclei. If we increase the resolution scale to resolve quarks and gluons, we can use Lattice QCD to study nuclear interactions [137, 138] and currents [139, 140]. However, the method is not yet operational for accurate predictions of atomic nuclei [21-25] at physical quark masses. Nevertheless, short of practical and computational challenges, some of the pion-nucleon couplings of χEFT have been computed on the lattice [141]. Although still operating at unphysical pion masses, lattice results can be extrapolated in the infrared using EFT methods [142]. This extrapolative approach has turned out to be particularly valuable in the data-scarce hyperon sector [143] to, e.g., elucidate the role of strangeness in dense nuclear matter [144]. If we instead decrease the resolution scale, likely at the cost of predictive power, we can integrate out the pion to obtain a systematically improvable pionless EFT [145, 146] for which we can solve the Schrödinger equation and perform *ab initio* computations of processes at very low external momenta. Continuing in this direction, one can devise halo EFT [147, 148], and EFTs for collective phenomena [149–153]. Although these latter two methods are systematically improvable, i.e., they are equipped with a power-counting scheme, they have even less predictive power because they exhibit a relatively small breakdown scale and are tailored to analyze a particular class of low-energy phenomena.

The traditional shell model can be formulated as an *ab initio* approach if one derives the valence-space interaction from a few-nucleon Hamiltonian, based on χEFT, using a systematically improvable prescription [52, 154–156]. Likewise, coarse-grained representations of nuclear phenomena, like those provided by density functional theory, might be cast as an *ab initio* method one day if we can link them to low-energy interactions derived from χEFT [157, 158]. However, this has not yet come to fruition [159, 160].

Finally, it is essential to point out that different and sometimes conflicting assumptions regarding the power counting scheme and its meaning are in use. Besides the foundations, which are covered at length in, e.g., Refs. [79, 161–165], all power counting schemes strive to furnish an EFT description of the nuclear interaction that become increasingly refined at higher orders of the expansion. To test this, we must perform calculations to predict nuclear observables. This is an important example of how the nuclear *ab initio* method and nuclear EFTs are intimately connected and how they can benefit from each other.

# 6 Summary

The *ab initio* method should not be confused with nuclear EFT. The *ab initio* method includes the ideas of EFT in the sense that it is systematically improvable, and one starts from degrees of freedom determined by the relevant scale separation and resolution. However, the *ab initio* method is also something more. What this "more" is, has not been specified or discussed much in our community. Naturally, misunderstandings and controversies often arise, and one may meet questions like: "Is this really *ab initio*?" With this paper we hope to bring some clarity to that question.

In our view, the *ab initio* method should set *the beginning* at a resolution scale that maximizes the method's predictive power and enables reliable predictions for phenomena at multiple energy scales ranging from a few tens of keV's to hundreds of MeV's. This implies that nucleons are currently the appropriate degrees of freedom for the *ab initio* method. However, it is an open question whether the beginning can be shifted to an even finer resolution scale, e.g., quarks and gluons while increasing the predictive power across energy scales significantly. We interpret the *ab initio* method as a systematically improvable approach employing Lagrangians, Hamiltonians, or energy density functionals derived from the Standard Model according to the principles of EFT. Subsequently solving for observables using numerically exact methods or, if necessary, controlled approximations that allow for systematic predictions with quantified uncertainties.

Despite our best efforts, tensions between theoretical predictions and experimental results remain. It is however clear that the *ab initio* method offers a unique advantage for estimating the uncertainties necessary for assessing the significance of discrepancies between theory and experiment.

# Author contributions

All authors listed have made a substantial, direct, and intellectual contribution to the work and approved it for publication.

# Funding

This work was supported by the European Research Council (ERC) under the European Unions Horizon 2020 research and innovation program (Grant agreement No. 758027), the Swedish Research Council (Grants Nos. 2017-04234, 2020-05127, and 2021-04507), the U.S. Department of Energy, Office of Science, Office of Nuclear Physics, under Award Nos. DE-FG02-96ER40963 and DE-SC0018223. This research used resources of the Oak Ridge Leadership Computing Facility at the Oak Ridge National Laboratory, which is supported by the Office of Science of the U.S. Department of Energy under Contract No. DE-AC05-00OR22725.

# Conflict of interest

The authors declare that the research was conducted in the absence of any commercial or financial relationships that could be construed as a potential conflict of interest.

# Publisher's note

All claims expressed in this article are solely those of the authors and do not necessarily represent those of their affiliated organizations, or those of the publisher, the editors and the reviewers. Any product that may be evaluated in this article, or claim that may be made by its manufacturer, is not guaranteed or endorsed by the publisher.

Ekström et al.	10.3389/fphy.2023.1129094